# A Review of Deep Learning Techniques for Protein Function Prediction

Divyanshu Aggarwal
*Dept of Biotechnology,*
*Delhi Technological University,*
New Delhi, India
divyanshuggrwl@gmail.com

Yasha Hasija
*Dept of Biotechnology,*
*Delhi Technological University,*
New Delhi, India
yashahasija06@gmail.com (Corresponding author)

*Abstract*—Deep Learning and big data have shown tremendous success in bioinformatics and computational biology in recent years; artificial intelligence methods have also significantly contributed in the task of protein function classification. This review paper analyzes the recent developments in approaches for the task of predicting protein function using deep learning. We explain the importance of determining the protein function and why automating the following task is crucial. Then, after reviewing the widely used deep learning techniques for this task, we continue our review and highlight the emergence of the modern State of The Art (SOTA) deep learning models which have achieved groundbreaking results in the field of computer vision, natural language processing and multi-modal learning in the last few years. We hope that this review will provide a broad view of the current role and advances of deep learning in biological sciences, especially in predicting protein function tasks and encourage new researchers to contribute to this area.

## I. INTRODUCTION

Proteins play a large role in the cellular machinery of the living organism. It is very important to know the function protein while conducting any proteomic research on that particular protein. However, more than 40% of the NCBI database's protein sequences have no assigned function as of 2013 [1]. This can be induced by the fact that it is expensive, high in processing time, and challenging to determine a protein's function using functional assays. [1] This arises the need for a computational method to determine a protein's function from the raw data obtained from the high throughput techniques, including but not limited to protein sequence, protein structure, gene expression profile and protein-protein interaction data.

Traditional approaches to classification for protein function attempt to identify the evolutionary relationship between a new protein and a query protein [1]. A high sequence similarity score can suggest a high probability of 2 proteins originating from a single evolutionary source [1]. However, it is well established that proteins with high sequence alignment or sequence similarity score may or may not show similar functions.[2] Such erroneous annotations may lead to propagation and amplification in large databases quickly.

Such errors are not only due to basic transfer strategies based on homology, but also due to the manual data analysis

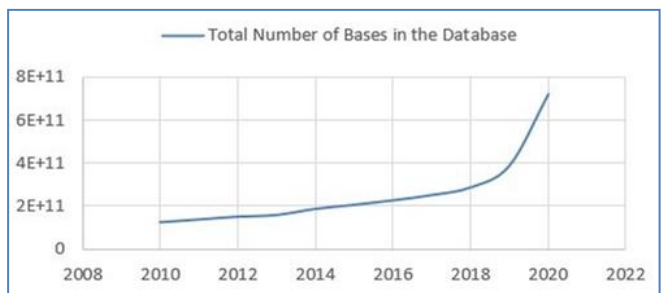

Graph 1. Growth of Gene Bank [3]

Process. Such problems have also been tried to tackle using different data, including but not limited to 3D structure similarity, gene expression profile similarity and genomic expression profile similarity.

In the past few years, the growth of biological databases and significant advancements in computing resources creates an opportunity for the scientific community to build novel deep learning models and architectures to tackle such problems. As the latest trend in recent developments in natural language processing research and sequential data preprocessing using deep neural networks (DNN), convolutional neural networks (CNN), recurrent neural networks (RNN), long- term and short-term memory (LSTM), and attention-based transformer models, it is important to evaluate the suitability, feasibility, sustainability and explain ability of such models for the given task. In the sense of protein function classification, this analysis aims to provide an overview of these techniques.

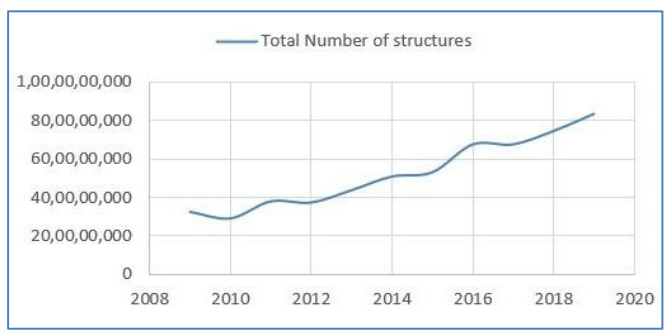

Graph 2. Growth of Protein Data Bank [4]

This paper outlines the latest trends in the development of automated annotation of protein function from raw data using deep learning. The flow of paper will be as follows; first, we discuss about the approached for prediction protein function from raw without using deep learning. (section 2). Next, we describe the recent deep learning and machine learning advances and how they are used in the protein function classification perspective (Section 3). We discuss related work and concluding remarks in the section after that (Section 4 and 5, respectively).

## II. CLASSICAL COMPUTATIONAL TECHNIQUES FOR PROTEIN FUNCTION PREDICTION

Understanding how the cell functions enables one to learn how the protein works. Usually, a protein function is specified as the molecular function of the protein; material transport, gene regulation, catalysis of biochemical reactions (enzymes) and others are some protein functions. The protein functions in many ways; it can interact with other proteins in order to perform its roles biologically. The mutation of the amino acid sequence may result in certain diseases from a phenotypical perspective. Factors like, but not limited to, such as, the environment of the cell, for example, temperature, subcellular protein position, may also influence the role of a protein. For standardizing functional annotation and enabling computer processing to explain various aspects of protein functions, several classification schemes have suggested specific vocabulary. Gene Ontology is the most generally recognized (GO). Three aspects of protein functions are defined by GO: molecular structure, biological mechanism, and cellular positioning. An ontology is constructed as a directed acyclic graph, here vertices reflect function of protein and their relationships are represented by the edges. For their practical predictions from this machine-readable vocabulary rather than natural language style, methods of prediction using computation should have a uniform, output, which can show errors within the annotation level. [5], [6], [7]

The most common means of identifying a protein's function by using the already annotated sequences is through BLAST. The query sequence is matched with the sequences from the database, and the similarity score is used to estimate the function of the protein with different sequences. BLAST uses heuristic algorithms to match strings. The genes with high similarity scores are said to be homologs. It is assumed that protein sequences with similar homology have a similar function, however, sequence matching is not the most accurate way of determining the protein's function; 2 highly similar protein sequences are not supposed to have same Gene Ontology in all the scenarios [8].

Gene fusion and phylogenetic profile are another 2 features which can give good accuracy in the task of prediction protein function. Gene neighborhood strategies use the idea that proteins whose translated genes are located near each other on a chromosome are deemed to be linked in functional way in different genomes. In one genome that is fused into another genome to form a single gene, the idea that group of genes are required focuses on gene fusion-based approaches. Another ground-breaking method for predicting functional relationships is to compare the phylogenetic profiles of proteins. A collection of bits that indicate a homologous presence or absence in each genome is known as protein phylogenetic profile. Similar profiling may also imply that gene which has produced the proteins have evolved together and may indicate function conservation. [9], [10]

It is also possible to use motifs and domain details. A domain is a part of the amino acid sequence of proteins that folds independently of the rest of the structure into a stable structure. A motif is a very short stretch of sequences of amino acids that theoretically encodes the structure of the protein. These are used for the metadata of protein, such as length of sequence, composition of amino acids, and physicochemical properties of a sequence. [2]

However, the structure of the protein is the most informative aspect of a protein to describe the function of a protein. Due to the Structural Genomics Initiative, we today have a database called protein data bank, which contains the structure of millions of protein sequences. The query protein is structurally matched. The protein sequences are matched in 2 ways, globally and locally. In global alignment, the entire structure is matched, which in local structural alignment, local regions like motifs and domains are matched since they are a better criterion for identifying the protein's function. The motifs and domains can capture the intramolecular interactions of a protein, while global alignment methods are able to identify the intermolecular interactions. Intermolecular interactions are generally performed on the surface, which are responsible for a protein's biochemical properties. [11]

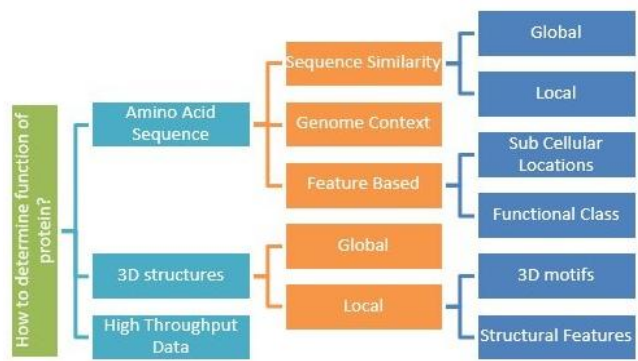

Fig. 1.  .

With the emergence of high-throughput protein sequencing technologies, vast quantities of data have been made available in biological gene expression and protein-protein interaction databases. The gene expression is measured by the amount of the protein produced by a gene in specific conditions that performs a particular gene. It is probable that proteins whose genes are co-expressed have the same role. Proteins subsequently perform a particular role by communicating with other proteins; hence similar protein-protein interactions can lead to similar protein functions.

## III. USING DEEP LEARNING FOR PROTEIN FUNCTION PREDICTION

Deep learning has achieved significant momentum in the past few years with the advancements in the computing power of the central processing units and graphical processing units and the rapid expansion of biological databases. Deep learning algorithms are capable of extracting hidden features and patterns from the data are also able to form non-linear classification boundaries. [12], [13] Deep learning also scales well with the growth in data volumes, and its application to computer vision and natural language processing can be adapted to the structure of 3D proteins and protein sequences, respectively. With the recent advances in neural graph networks, they can also be applied in protein-protein interaction and phylogenetic databases.

*A. Data Sources*

There are multiple data sources from where the data can be obtained to train the model from such tasks. Uniprot KB and PDB remains to be the open databases that can be freely accessed. [14], [15] The data from such databases can be downloaded in bulk and can be cleaned and preprocessed to train such models. However, the CAFA dataset released by The Function Special Interest Group is a dataset mainly meant for the challenge of building computational models for protein function classification. [16] However some models have also used STRING db in the past.

*B. Deep Learning Paradigm*

This paper presents the analysis of modern deep learning techniques in the context of protein function classification. The classification of the function of a protein can be classified as a supervised machine learning task. Since neural networks in general settings cannot use unsupervised learning tasks, we cannot use clustering techniques or other unsupervised techniques. Deep learning research has gathered momentum in the past few years. This can be attributed to the advancements in modern machines' computational capacity and availability of large publicly available databases. The sector has had a major influence, including, to name a few, banking, healthcare, social media. Recent developments in the processing of natural language and computer vision by deep learning can have an immense impact on genomics and proteomics research. [13]

Recent developments in algorithm development, such as DNNs, RNNs, CNNs, LSTMs, and attention models for image and text data, the State of the art and near-human precision have been achieved already, we can only wonder how they can do in such biological tasks where the enormous amounts of similar data are accessible. However, the challenge of data representation arises while solving such tasks using deep learn- ing. While there are pre-trained word embeddings available for natural language, it is difficult to say so for biological sequence databases. However, researchers have provided some protein embedding in recent years, and the quality of such models is not on par with natural language models, trained with similar volumes of data. The reason might be the relatively small vocabulary size of protein sequences with 20 amino acids known, whereas more than 100,000 words are taken in the vocabulary while training a natural language model. [17] The subsequent sections will explore the recent advancements in deep learning and their implication in the protein classification task.

*C. Protein Function prediction using sequence only*

With the advances in text classification tasks using deep learning, one can only get curious about how these techniques can be applied to biological sequence classification. Related biological problems have been further addressed by developments in word embedding methods, 1D CNN, FastText word embedding, DNNs, and RNNs by adapting methods from natural language processing tasks. We will address some of the methods mentioned in this subsection to predict a protein's function solely from raw sequence data.

*1) Combination of 1D Convolutional Neural Network and Deep Neural Networks:* Convolutional Neural Networks are the deep learning algorithms capable of extracting unique geographic features from the data. They are also used to reduce the data's size by amplifying and representing localized features in the final representation. The convolutional Kernel translates over an image or text and captures the localized information during translation. A 1D Convolutional Neural Network convolves through a protein sequence representation in 1 dimension. [18], [19], [20]

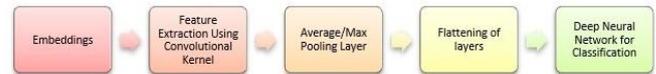

Fig. 2. Illustration of CNN and DNN for Function Prediction

The convolutional Kernel is translated over the word embedding of the sequence, concatenated word byword for a sequence. There are several alternatives available for natural languages like FastText or word2vec, while for protein sequences, ProtVec or prot2vec is a popular choice. [21], [22], [23]

*2) Using Recurrent Neural Networks:* Recurrent neural networks are neural networks that can identify the temporal characteristics of the sequence, i.e., a network graph is a graph where a directed graph forms connection between nodes along a time sequence. There are dynamic networks. They are useful in learning features from variable length features. This makes them useful for processing variable-length sequences in the input. Such models can also be used for sequence-to-sequence tasks like machine translation.

The modified version of RNN is known as Long Term Short Memory (LSTM). LSTMs are capable of storing particular temporal features in the hidden State of the cell. [24] This makes it more accurate for longer sequences and sequences with high length variability. The LSTM also tackles the problem of vanishing and exploding gradient problem better than vanilla RNNs. [12], [24], [25]

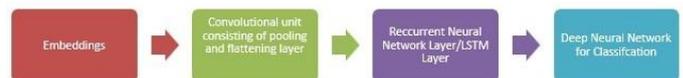

Fig. 3. Schematic Diagram of protein function classification using Recurrent Neural Networks

Bi-LSTMs, as the name suggests, are bidirectional LSTM neural network architectures. These can capture the bidi-rectional features of the sequence. Hence, they can have advantages in tasks with bidirectional nature like sequence classification. These models can be used by feeding embedding layers directly to the LSTM models, where they extract out the sequential features, and they can be concatenated or added further. The resulting vector is then fed to a feed-forward layer or multiple feed-forward layers, which can then be passed through a SoftMax layer for the classification task.

*3) Transformers for Protein Sequence Classification:* With the introduction of attention-based models in 2017 by google research, natural language processing has changed drastically. The attention-based transformers can process long sequences much better than LSTMs and can utilize the parallel GPU architecture for faster and more cost-effective training. The Transformer models can extract out location-specific features and give weights to tokens based on

context relevance. [26]

The transformer models scale up well with the increasing data and the increasing number of parameters. With GPT and BERT's introduction, large language models are becoming more and more mainstream with time. [27], [28] The models can grasp the link between sequence tokens very well and are the present State of the art for most NLP assignments. Moreover, these models can also form word embeddings through self-supervised tasks like masked language modeling and next-word prediction. [17] These representations can then be used to for finetuning for the number of other mainstream tasks. Such self-supervised learning techniques can also be used for protein sequences as demonstrated by protBERT and proBERTa. [29], [30]

### D. Multi-modal Deep Learning for Protein Function Classifi-cation

In past few years, multi-modal methods of deep learning have gained huge traction. A basic multi-modal deep learning technique can be performed by combining the image features and text features. The researchers have previously used sequence data, genomic expression, 3D structures, and data on protein- protein interaction to predict the activity of the protein. With the advances in Natural Language Processing and Computer Vision, such techniques seem promising for the Particular Task. The Machines can learn in cognizance with different data representations to make predictions. The sequence data can provide information about the amino acid sequences and their positioning, while 3D structure can show structural domains and motifs, which are a significant indicator of the protein's function. The genomic expression data can capture the patterns in the genomic expression and the corresponding GO annotation, while protein-protein interaction databases are useful for finding the patterns in protein-to-protein interactions. [31], [32], [33], [34]

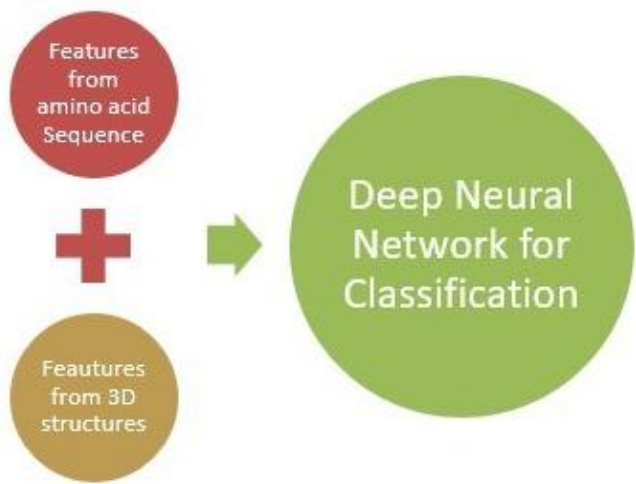

Fig. 4. Multi-modal Deep Learning Illustration.

The Multi-modal deep learning techniques, given the high compute power available these days, seems promising for such tasks. With the increase in high throughput technologies, the database for such features is only expected to grow further, providing us with more data for training these models. There is also a significant momentum gained in multi-modal deep learning research, which will further drive such models' accuracy.

### E. Using Autoencoders

A few researchers in the past have also tried using denoising autoencoders for extracting features from the sequence and structure alike. [31], [35], [36], [37] These features are then fed into any other classifier of choice. This denoising autoencoder is trained to produce denoised outputs that are generated randomly with some average and standard deviation according to a normal distribution. However, the technique is not just limited to multi-modal techniques. Masked language modeling is shown in protBERT, and ProtBERTa is also denoising autoencoder task applied on raw protein sequence only. [29], [30]

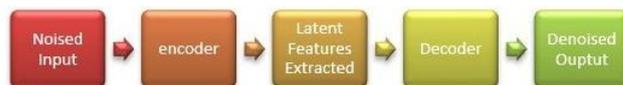

Fig. 5. Autoencoder model architecture.

## IV. RELATED WORK

We reviewed 9 recent models built by researchers around the world for this task. We reported AUC score of each model with respect to the CAFA challenge and human and yeast STRING networks metric. [16], [38] We have reported the methodologies and the resulting Fmax score in table 1.

According to **figure 6,** the 1D CNN combined with DNN seems to be the clear preference for function prediction task. The models are simple and perform well in the sequence classification tasks. At the same time, the transformer models seem to be least explored for this task.

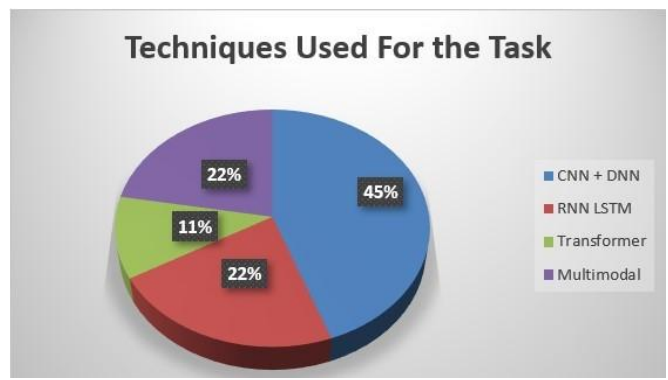

Fig. 6. Distribution of all the techniques used by the model reviewed models

From the above data we can infer that simple techniques like 1D CNN + DNN perform well on the raw sequence data. They can be improved by combining with PPI data; gene expression data are alike. Around the same time, SOTA NLP models such as transformers are not on par with other conventional techniques. However, there is not much research that is done on transformer models for such downstream tasks. It shall be interesting to see the implications of such models further in the future for such downstream tasks, especially the use of protBERT and PRoBERTa pre-trained models finetuned for protein function classification. [29], [30]

TABLE I. RECENT MODELS AND THEIR PERFORMANCE IN THE PROTEIN FUNCTION CLASSIFICATION

| Model | Authors | Data Published | Methods Used | Fmax | Validation Dataset |
|---|---|---|---|---|---|
| DeepGO | Kulmanov et al. | February 2018 | CNN+DNN | 0.47 | CAFA3 |
| DeepNF | Gligorijevic et al. | June 2018 | Multi-modal Deep Learning | 0.42 | STRING |
| DeepPred | Rifaioglu et al. | May 2019 | Hierarchical DNN | 0.50 | CAFA3 |
| DeepGOPlus | Kulmanov et al. | July 2019 | CNN+DNN | 0.54 | CAFA3 |
| UDSMProt | Strodthoff et al. | January 2020 | LSTMs | 0.58 | CAFA3 |
| SDN2GO | Cai et al. | April 2020 | CNN+DNN | 0.56 | CAFA3 |
| MultiPredGO | Giri et al. | September 2020 | Multi-modal Deep Learning | 0.37 | CAFA3 |
| TALE+ | Cao et al. | September 2020 | Transformer | 0.67 | CAFA3 |

Fmax Scores are based on the Molecular Function Prediction of the models.

## V. CONCLUDING REMARKS

In the world of bioinformatics, developments in deep learn- ing have had a huge influence. The researchers can now make sophisticated deep learning Models that can derive concealed characteristics from raw biological data. This can remove the requirement of hand-crafted features for machine learning tasks. Moreover, such models can be scaled well with increasing data and compute capacity. [42] Thus, such models can be monumental in building an automated system for annotating protein functions through raw data.

It would be interesting in the future to see how some pre-trained models like prot-BERT [30] and PRoBERTa [43] performs when finetuned for the downstream tasks for protein function prediction. Such protein vector representations are said to be SOTA in the self-supervised task for masked language modelling. The recent ESM-1b protein language mode by Facebook AI Research is also another contender. [44] With recent advances in multi-modal deep learning and the production of multi-dimensional data in biological databases, the characteristics derived from various types of raw data, such as 3D structure, protein-protein interaction, are used to predict protein function, the gene expression profile can be paired with current SOTA protein sequence representations, which is likely to boost the Fmax scores on the CAFA challenge further.

## REFERENCES

[1] J. Bernardes and C. Pedreira, "A Review of Protein Function Prediction Under Machine Learning Perspective," Recent Patents on Biotechnology, vol. 7, no. 2, pp. 122–141, 2013. [Online]. Available: 10.2174/18722083113079990006;https://dx.doi.org/10.2174/18722083113079990006

[2] F. Eisenhaber, 2013.

[3] W. Genbank and Statistics, 2021. [Online]. Available: https://www.ncbi. nlm.nih.gov/genbank/statistics/.Accessed28

[4] 2021. [Online]. Available: http://www.wwpdb.org/stats/download. Accessed28

[5] M. Ashburner, C. A. Ball, J. A. Blake, D. Botstein, H. Butler, J. M. Cherry, A. P. Davis, K. Dolinski, S. S. Dwight, J. T. Eppig, M. A. Harris, D. P. Hill, L. Issel-Tarver, A. Kasarskis, S. Lewis, J. C. Matese, J. E. Richardson, M. Ringwald, G. M. Rubin, and G. Sherlock, "Gene Ontology: tool for the unification of biology," Nature Genetics, vol. 25, no. 1, pp. 25–29, 2000. [Online]. Available: 10.1038/75556;https://dx.doi.org/10.1038/75556

[6] A. D.Diehl, J. A. Lee, R. H.Scheuermann, and J. A. Blake, "Ontology development for biological systems: immunology," Bioinformatics, vol. 23, no. 7, pp. 913– 915, 2007. [Online]. Available: 10.1093/bioinformatics/btm029;https://dx.doi.org/10.1093/bioinformatics/btm029

[7] J. D. Osborne, J. Flatow, M. Holko, S. M. Lin, W. A. Kibbe, L. Zhu, M. I. Danila, G. Feng, and R. L. Chisholm, "Annotating the human genome with Disease Ontology," BMC Genomics, vol. 10, no. Suppl 1, pp. S6–S6, 2009. [Online]. Available: 10.1186/1471-2164-10-s1-s6;https://dx.doi.org/10.1186/1471-2164-10-s1-s6

[8] S. F. Altschul, W. Gish, W. Miller, E. W. Myers, and D. J. Lipman, "Basic local alignment search tool," Journal of Molecular Biology, vol. 215, no. 3, pp. 403–410, 1990. [Online]. Available: 10.1016/s0022- 2836(05)80360-2;https://dx.doi.org/10.1016/s0022-2836(05)80360-2

[9] J. Huerta-Cepas, D. Szklarczyk, K. Forslund, H. Cook, D. Heller, M. C. Walter, T. Rattei, D. R. Mende, S. Sunagawa, M. Kuhn, L. J. Jensen, C. von Mering, and P. Bork, "eggNOG 4.5: a hierarchical orthology framework with improved functional annotations for eukaryotic, prokaryotic and viral sequences," Nucleic Acids Research, vol. 44, no. D1, pp. D286–D293, 2016. [Online]. Available: 10.1093/nar/gkv1248;https://dx.doi.org/10.1093/nar/gkv1248

[10] B. E, "Genome-scale phylogenetic function annotation of large and diverse protein families," Genome Res, vol. 21, pp. 1969–1980.

[11] D. Cozzetto, F. Minneci, H. Currant, and D. T. Jones, "FFPred 3: feature-based function prediction for all Gene Ontology domains," Scientific Reports, vol. 6, no. 1, 2016. [Online]. Available: 10.1038/srep31865;https://dx.doi.org/10.1038/srep31865

[12] Y. LeCun, Y. Bengio, and G. Hinton, "Deep learning," Nature, vol. 521, no. 7553, pp. 436–444, 2015. [Online]. Available: 10.1038/nature14539;https://dx.doi.org/10.1038/nature14539

[13] C. Cao, F. Liu, H. Tan, D. Song, W. Shu, W. Li, Y. Zhou, X. Bo, and Z. Xie, "Deep Learning and Its Applications in Biomedicine," Proteomics Bioinforma, vol. 16, pp. 17–32, 2018.

[14] R. Apweiler, "The Universal Protein resource (UniProt)," Nucleic Acids Res, vol. 36, 2008.

[15] H. Berman, K. Henrick, and H. Nakamura, "Announcing the worldwide Protein Data Bank," Nature Structural & Molecular Biology, vol. 10, no. 12, pp. 980–980, 2003. [Online]. Available: 10.1038/nsb1203-980;https://dx.doi.org/10.1038/nsb1203-980

[16] A. N. Zhou, Y. Jiang, T. R. Bergquist, A. J. Lee, B. Z. Kacsoh, W. Crocker, K. A. Lewis, G. Georghiou, H. N. Nguyen, M. N. Hamid, L. Davis, T. Dogan, V. Atalay, A. S. Rifaioglu, A. Dalkiran, R. Cetin-Atalay, C. Zhang, R. L. Hurto, P. L. Freddolino, Y. Zhang, P. Bhat, F. Supek, J. M. Fernández, B. Gemovic, V. R. Perovic, R. S. Davidovic´, N. Sumonja, N. Veljkovic, E. Asgari, M. Mofrad, G. Profiti, C. Savojardo, P. L. Martelli, R. Casadio, F. Boecker, I. Kahanda, N. Thurlby, A. C. Mchardy, A. Renaux, R. Saidi, J. Gough, A. A. Freitas, M. Antczak, F. Fabris, M. N. Wass, J. Hou, J. Cheng, J. Hou, Z. Wang, A. E. Romero, A. Paccanaro, H. Yang, T. Goldberg, C. Zhao, L. Holm, P. Törönen, A. J. Medlar, E. Zosa, I. Borukhov, I. Novikov, A. Wilkins, O. Lichtarge, P. H. Chi, W. C. Tseng, M. Linial, P. W. Rose,C. Dessimoz, V. Vidulin, S. Dzeroski, I. Sillitoe, S. Das, J. G. Lees, C. T. Jones, C. Wan, D. Cozzetto, R. Fa, M. Torres, A. W. Vesztrocy, J. M. Rodriguez, M. L. Tress, M. Frasca, M. Notaro, G. Grossi, A. Petrini, M. Re, G. Valentini, M. Mesiti, D. B. Roche, J. Reeb, D. W. Ritchie, S. Aridhi, S. Z. Alborzi, M. D. Devignes, D. Koo, R. Bonneau, V. Gligorijevic´, M. Barot, H. Fang, S. Toppo, E. Lavezzo, M. Falda, M. Berselli, S. Tosatto, M. Carraro, D. Piovesan, H. U. Rehman, Q. Mao, S. Zhang, S. Vucetic, G. S. Black, J. D. Larsen, D. J. Omdahl, A. R. Sagers, L. W. Suh, E. Dayton, J. B. Mcguffin, L. J. Brackenridge, D. A. Babbitt, P. C. Yunes, J. M. Fontana, P. Zhang, F. Zhu, S. You, R. Zhang, Z. Dai, S. Yao, W. Tian, R. Cao, R. Chandler, C. Amezola, M. Johnson, D. Chang, J. M. Liao, H. W. Liu, Y. W. Pascarelli, S. Frank, Y. Hoehndorf, R. Kulmanov, M. Boudellioua, I. Politano, G, D. Carlo, S. Benso, A. Hakala, K. Ginter, F. Mehryary, F. Kaewphan, S. Björne, J. Moen, H. Tolvanen, M. Salakoski, T. Kihara, D.

Jain, A. Šmuc, T. Altenhoff, A. Ben-Hur, A. Rost, B. Brenner, S. E. Orengo, C. A. Jeffery, C. J. Bosco, G. Hogan, D. A. Martin, M. J. O'donovan, C. Mooney, S. D. Greene, C. S. Radivojac, P. Friedberg, and I, 2019. [Online]. Available: https://doi.org/10.1101/653105

[17] J. Vig, A. Madani, L. R. Varshney, C. Xiong, R. Socher, and N. F. Rajani, 2020.

[18] M. Kulmanov and R. Hoehndorf, "DeepGOPlus: Improved protein function prediction from sequence," Bioinformatics, vol. 36, pp. 422–429, 2020.

[19] M. Kulmanov, M. A. Khan, and R. Hoehndorf, "DeepGO: predicting protein functions from sequence and interactions using a deep ontology-aware classifier," Bioinformatics, vol. 34, no. 4, pp. 660–668, 2018. [Online]. Available: 10.1093/bioinformatics/btx624;https://dx.doi.org/10.1093/bioinformatics/btx624

[20] Y. Cai, J. Wang, and L. Deng, "SDN2GO: An Integrated Deep Learning Model for Protein Function Prediction," Frontiers in Bioengineering and Biotechnology, vol. 8, pp. 1–11, 2020. [Online]. Available: 10.3389/fbioe.2020.00391;https://dx.doi.org/10.3389/fbioe.2020.00391

[21] E. Asgari and M. R. K. Mofrad, "Continuous Distributed Representation of Biological Sequences for Deep Proteomics and Genomics," PLOS ONE, vol. 10, no. 11, pp. e0 141 287–e0 141 287, 2015. [Online]. Available: 10.1371/journal.pone.0141287;https://dx.doi.org/10.1371/journal.pone.0141287

[22] X. Lin, "DeepGS: Deep Representation Learning of Graphs and Sequences for Drug-Target Binding Affinity Prediction," Front Artif Intell Appl, vol. 325, pp. 1301–1308, 2020.

[23] 2021. [Online]. Available: https://github.com/yotamfr/prot2vec. Accessed28

[24] X. L. Liu, 2017. [Online]. Available: https://doi.org/10.1101/103994

[25] S. Hochreiter and J. Schmidhuber, "Long Short-Term Memory," Neural Computation, vol. 9, no. 8, pp. 1735–1780, 1997. [Online]. Available: 10.1162/neco.1997.9.8.1735;https://dx.doi.org/10.1162/neco.1997.9.8.1735

[26] A. Vaswani, N. Shazeer, N. Parmar, J. Uszkoreit, L. Jones, A. N. Gomez, Ł. Kaiser, and I. Polosukhin, "Attention is all you need," Advances in Neural Information Processing Systems. Neural information processing systems foundation, pp. 5999–6009, 2017.

[27] A. Radford, J. Wu, R. Child, D. Luan, and D. Amodei.

[28] J. Devlin, M. W. Chang, K. Lee, K. T. Google, A. I. Language, and Bert.

[29] S. C. Huang, L. Martinez-Nunez, R. T. Rupani, H. Platé, M. Niranjan, M. Chambers, R. C. Howarth, P. H. Sanchez-Elsner, T. Azodi, C. B. Tang, J. Shiu, S. H. Senior, A. W. Evans, R. Jumper, J. Kirkpatrick, J. Sifre, L. Green, T. Qin, C. Žídek, A. Nelson, A. Bridgland, A. Pene-dones, H. Petersen, S. Simonyan, K. Crossan, S. Kohli, P. Jones, D. T. Silver, D. Kavukcuoglu, K. Hassabis, D. Vig, J. Madani, A. Varsh-ney, L. R. Xiong, C. Socher, R. Rajani, N. F. Rives, A. Goyal, S. Meier, J. Guo, D. Ott, M. Zitnick, C. L. Ma, J. Fergus, R. Elnaggar, A. Heinzinger, M. Dallago, C. Rost, B. Shanehsazzadeh, A. Belanger, D. Dohan, D. Only, U. Madani, A. Mccann, B. Naik, N. Keskar, N. S. Anand, N. Eguchi, R. R. Huang, P. S. Socher, R. Lu, A. X. Zhang, H. Ghassemi, M. Moses, A. Nambiar, A. Liu, S. Heflin, M. Maslov, S. Hopkins, and M, pp. 1–11.

[30] A. Elnaggar, M. Heinzinger, C. Dallago, G. Rehawi, Y. Wang, L. Jones, Gibbs, T. Feher, C. Angerer, M. Steinegger, D. Bhowmik, B. Rost, and Prottrans.

[31] V. Gligorijevic', M. Barot, and R. Bonneau, "deepNF: deep network fusion for protein function prediction," Bioinformatics, vol. 34, no. 22, pp. 3873–3881, 2018. [Online]. Available: 10.1093/bioinformatics/bty440;https://dx.doi.org/10.1093/bioinformatics/bty440

[32] T. Baltrušaitis, C. Ahuja, L. P. Morency, and Multimodal.

[33] S. J. Giri, P. Dutta, P. Halani, and S. Saha, "MultiPredGO: Deep Multi-Modal Protein Function Prediction by Amalgamating Protein Structure, Sequence, and Interaction," IEEE Journal of Biomedical and Health Informatics, vol. 2194, pp. 1–1, 2020. [Online]. Available: 10.1109/jbhi.2020.3022806;https://dx.doi.org/10.1109/jbhi.2020.3022806

[34] Q. Mcnamara, D. L. Vega, A. Yarkoni, and T, "Developing a comprehensive framework for multi-modal feature extraction," Proceedings of the ACM SIGKDD International Conference on Knowledge Discovery and Data Mining, pp. 1567–1574, 2017.

[35] R. Cao, C. Freitas, L. Chan, M. Sun, H. Jiang, and Z. Chen, "ProLanGO: Protein Function Prediction Using Neural Machine Translation Based on a Recurrent Neural Network," Molecules, vol. 22, no. 10, pp. 1732–1732, 2017. [Online]. Available: 10.3390/molecules22101732;https://dx.doi.org/10.3390/molecules22101732

[36] R. Bonetta and G. Valentino, "Machine learning techniques for protein function prediction," Proteins: Structure, Function, and Bioinformatics, vol. 88, no. 3, pp. 397–413, 2020. [Online]. Available: 10.1002/prot.25832;https://dx.doi.org/10.1002/prot.25832

[37] Q. Meng, D. Catchpoole, D. Skillicorn, and P. J. Kennedy, Relational Autoencoder for Feature Extraction.

[38] D. Szklarczyk, A. L. Gable, D. Lyon, A. Junge, S. Wyder, J. Huerta-Cepas, M. Simonovic, N. T. Doncheva, J. H. Morris, P. Bork, L. J. Jensen, and C. von Mering, "STRING v11: protein–protein association networks with increased coverage, supporting functional discovery in genome-wide experimental datasets," Nucleic Acids Research, vol. 47, no. D1, pp. D607–D613, 2019. [Online]. Available: 10.1093/nar/gky1131;https://dx.doi.org/10.1093/nar/gky1131

[39] A. S. Rifaioglu, T. Doğan, M. J. Martin, R. Cetin-Atalay, and V. Atalay, "DEEPred: Automated Protein Function Prediction with Multi-task Feed-forward Deep Neural Networks," Scientific Reports, vol. 9, no. 1, pp. 1–16, 2019. [Online]. Available: 10.1038/s41598-019-43708-3;https://dx.doi.org/10.1038/s41598-019-43708-3

[40] N. Strodthoff, P. Wagner, M. Wenzel, and W. Samek, 2019. [Online]. Available: https://doi.org/10.1101/704874

[41] Y. Cao and Y. Shen, 2020. [Online]. Available: https://doi.org/10.1101/2020.09.27.315937

[42] J. Hestness, S. Narang, N. Ardalani, G. Diamos, H. Jun, H. Kianinejad, Patwary, Y. Yang, and Y. Zhou, 2017.

[43] A. Nambiar, S. Liu, M. Hopkins, M. Heflin, S. Maslov, and A. Ritz, 2020.

[44] A. Rives, S. Goyal, J. Meier, D. Guo, M. Ott, C. L. Zitnick, J. Ma, and R. Fergus, pp. 622 803–622 803, 2019.